\documentclass[preprint,12pt]{elsarticle}
\usepackage{amssymb}
\usepackage{graphicx}% Include figure files
\usepackage{dcolumn}% Align table columns on decimal point
\usepackage{bm}% bold math

\journal{Nuclear Physics A}
\begin{document}

\begin{frontmatter}

\title{\bf Pion mass effects on axion emission from neutron stars through NN bremsstrahlung processes}

\author{S. Stoica}
\address{Horia Hulubei National Institute of Physics and Nuclear Engineering, P.O. Box MG-6, 76900 Bucharest-Magurele, Romania }
\address{Horia Hulubei Foundation, Atomistilor 407, Bucharest-Magurele, Romania}
\ead{stoica@theory.nipne.ro}
\author{B. Pastrav}
\address{Horia Hulubei National Institute of Physics and Nuclear Engineering, P.O. Box MG-6, 76900 Bucharest-Magurele, Romania }
\ead{bpastrav@theory.nipne.ro}
\author{J. E. Horvath}
\address{Instituto de Astronomia, Geofisica e Ciencias Atmosfericas,
 Universidade de S\~ao Paulo, Rua do Matao, 1226, 05508-900, Cidade Universitaria S\~ao Paulo SP, Brazil}
\ead{foton@astro.iag.usp.br}
\author{M. P. Allen}
\address{CEFET-SP, R. Pedro Vicente, 625 01109-010, Caninde, S\~ao Paulo SP, Brazil}

\begin{abstract}
The rates of axion emission by nucleon-nucleon bremsstrahlung are
calculated with the inclusion of the full momentum contribution
from a nuclear one pion exchange (OPE) potential. The
contributions of the neutron-neutron (nn), proton-proton (pp) and
neutron-proton (np) processes in both the non-degenerate and
degenerate limits are explicitly given. We find that the
finite-momentum corrections to the emissivities are quantitatively
significant for the non-degenerate regime and temperature-dependent, and should affect the
existing axion mass bounds. The trend of these nuclear effects is
to diminish the emissivities.
\end{abstract}

\begin{keyword}
 nuclear effects \sep bremsstrahlung \sep axions \sep neutron stars
\PACS 14.80.Mz \sep 97.60.Jd \sep 95.30Cq

\end{keyword}

\end{frontmatter}

\section{Introduction}

The search for new particles/interactions beyond the Standard
Model is one of the most important tasks of particle physics.
While several candidates and proposals may be considered as
``exotic", in the sense of not being required by the data, it is
generally agreed that there are some possibilities definitely
expected as minimal extensions of it. Axions (\cite{[KIM79]}-\cite{[KIM87]}) belong to the
latter category, as expected from the Peccei-Quinn (\cite{[FWK78]}-\cite{[PCQ77]}) solution to
the strong CP-problem. Axions are pseudo Nambu-Goldstone boson
associated with the spontaneous breaking of the Peccei-Quinn
symmetry. Their masses and couplings are directly related to this
symmetry-breaking scale. Viable versions of axionic models
include the KSVZ (\cite{[KIM79]},\cite{[SVZ80]})
and DFSZ (\cite{[DFS81]}-\cite{[AZH80]}) axions coupled to hadrons only and to
leptons and hadrons respectively.

In addition to experimental efforts for a direct detection of
axions, astrophysical and cosmological arguments have played a key
role in their search. Actually, stringent bounds have been
obtained from the consideration of horizontal branch stars (\cite{[IWA84]}-\cite{[GRF95]}),
white dwarf cooling and SN1987A neutrino pulse duration (\cite{[YAK00]}),
among others. A general review of these arguments has been given
in \cite{[HGR98]} (see also \cite{[SIG97]}-\cite{[HPR01]} for a thorough account).

One of the main ingredients for an accurate calculation of axion
mass bounds is the emissivity in the nucleon bremsstrahlung
reaction $NN \rightarrow \, NNa$, thought to be dominant in
important astrophysical events, such as newly born neutron stars.
Calculations and discussions on the applicability of the
emissivity formulae were given in
\cite{[AZH80]},\cite{[YAK00]}-\cite{[BKT88]}. Quite generally, in these papers, the
calculations were performed for a one-pion exchange free nucleon
gas, leaving aside important issues later incorporated and
assessed, such as the effects of correlations between nucleons (see for instance \cite{[GRS95]}).
Attempts to link the emissivity to laboratory data have been also
made. We would like to present in this communication a
reassessment of the bremsstrahlung emissivity including the full
momentum dependence of the matrix elements. We found that the
hitherto neglected dependence produces large temperature-dependent
corrections to the rates independently of the many body effects, a
feature that points by itself to a revision of some of the
astrophysical bounds.

Axion emission are important for the evolution of stars
particularly for (hot) neutron stars (NS). For the conditions
relevant to the core of hot NS just after their formation ($T\sim
30-60$ MeV, $\rho \geq \rho_{0}$ (with $\rho_{0} \equiv 2.7 \times
10^{14} g cm^{-3}$ the nuclear matter density) the dominant
emission process are the nucleon-nucleon (NN) bremsstrahlung
(\cite{[BKT88]})

$$ n + n \rightarrow n + n + a  \eqno (a)$$
$$ p + p \rightarrow p + p + a \eqno (b)$$
$$ n + p \rightarrow n + p + a \eqno (c)$$

Previous calculations of the axion emissivities have
been performed by Iwamoto (\cite{[IWA84]},\cite{[IWA01]}) in the
degenerate (D) limit. For the NN interaction he used a OPE
potential in the Born approximation and found the expressions of
the energy-loss rates for all the processes above.

Later on, Brinkmann and Turner \cite{[BKT88]} calculated the axion
emission rates in the nondegenerate (ND) limit and for a general
degeneracy, for all three processes (a -c). They also could
check the result of Iwamoto for the equal-nucleon cases in the (D)
limit. For the NN interaction they used, however, constant nuclear
matrix elements. The same results for the ND limit was also
obtained previously by Turner in \cite{[TUR88]}. They reached the
conclusion that ND regime is a better approximation of
the axion emissivities for the conditions characteristic for a
newly born neutron star. Other calculations considering also a OPE potential where performed by Raffelt and Seckel \cite{[GRS95]}. They studied the axion emission rates of $NN\rightarrow NNa$ processes in order to determine their $S_{A}(\omega)$ structure function for NN interactions in neutron stars. With their calculations they concluded that the inclusion of pion mass effects do not reduce the axion emissivities by more than 50\% even for ND regime. Our results will show that the contributions of pion mass to the coresponding emissivities due to nuclear effects are temperature dependent, and for a certain temperature interval are larger than this, as we will see later.
Therefore, in this article we pay special
attention to the ND regime but we present also the results for D
regime. For the contribution of the $np$ process we consider
different chemical potentials for neutrons and protons, and as a
result we can span different degeneracy degrees for the two
species.

Analogously to the case of neutrino pair emission
(\cite{[BTH02]}-\cite{[FMX79]}), one
of the main difficulties for the calculation of axion emissivities
is the appropriate treatment of the strong NN interaction. In Iwamoto's calculations
for the D limit the effects were included by replacing the nucleon
momenta by their Fermi values in the angular part of the
phase-space integrals.

In the present work we include the full dependence on nuclear
momenta of the nuclear matrix elements (NME) in the calculations
of the axion emission rates by the NN bremsstrahlung processes (a
- c). Our results for $\epsilon_{aNN}$ separate explicitly the
part corresponding to constant NME, which corresponds to the
high-momentum limit of the previous works
(\cite{[IWA84]},\cite{[YAK00]}-\cite{[TBH20]}), from the part
including the missing nuclear effects due to the nucleon momenta
dependence of the NME to facilitate the comparisons
and further applications. We compare our results with those obtained
by Brinkmann and Turner (\cite{[BKT88]}) for the ND limit, while for the degenerate
regime we compare with those obtained in Ref. (\cite{[IWA84]}) and (\cite{[BKT88]}).

\section{Calculations}

The axion emission rate by NN bremsstrahlung is given by Fermi's
Golden Rule formula (see for instance \cite{[BKT88]})

$$ \epsilon_{aNN} = (2\pi)^4 {\int \left[\Pi_1^4 \frac{d^3{\bf p}_i}
{(2\pi)^3 2 E_i}\right] \frac{ d^3{\bf p}_{a}}{(2 \pi)^3 2E_a} E_a \left({\ S}\times \Sigma {\vert { M} \vert}^2 \right)
\delta^4(P){\it F(f)}} \eqno(1)$$

\noindent where ${\it F(f)} = {\it f_1 f_2 (1-f_3) (1-f_4)}$ is
the product of Fermi-Dirac distribution functions of the initial
(1,2) and final (3,4) nucleons, $ {\it f_i} =
\left(\exp^{\frac{E_i - \mu_i}{T}} + 1 \right)^{-1}$. In Eq. (1)
${\bf p}_i$ and $E_i$ (i=1,4) are the nucleon momenta and
energies, while ${\bf p}_a$ and $E_a$ are the corresponding axion
quantities; {\sl S} is a symmetry factor taking into account the
identity of the particles (1/4 for $nn$ and $pp$ channels and 1
for the $np$ channel) and $\mu_i$ are the chemical potentials of
the nucleons.

In the non-relativistic limit $E_i \sim m + \frac{{\bf
p}_i^2}{2m}$. Using the non-dimensional quantities \cite{[BKT88]}
$y = \hat{\mu}/T$ ($\hat{\mu}=\mu-m$) and $u_i = {\bf p}_i^2 / 2mT$, the  expressions
of the Fermi-Dirac functions read ${\it f_i} = \left(\exp^{u_i -
y_i} + 1\right)^{-1}$. The degenerate (D) limit satisfies $y
>> 1$, while in the non-degenerate (ND) limit $y << - 1$. For
${ S}\times \Sigma{\vert {M} \vert}^2$ we use  the
following expressions for the nuclear matrix elements (in the OPEP approximation):

$$ { S}\times \Sigma {\vert {M}\vert^2} = { S}\times \frac{256}{3}\cdot
g_{ai}^2 m^2 \left(\frac{f}{m_{\pi}}\right)^4 \cdot M_{NN} \eqno(2) $$

where $$g_{ai}=Cm_{N}/(f_{a}/N)$$ $C$ is a dimensionless factor of order unity, which is model dependent,
 $$ m_{n}\simeq m_{p}=940 MeV=m_{N} \rightarrow g_{an} \simeq g_{ap}=g_{ai}=C\cdot5.64\cdot10^{-10}$$ ($f_{a}=10^{10}GeV$ is the Peccei-Quinn symmetry breaking scale, $N=6$ represents the color anomaly of the Peccei-Quinn symmetry).
%We present our numerical results and our comparisons with Brinkmann and Turner's results by dividing the emissivities in both cases with $C^{2}$.If one wants to include the numerical results presented in this paper in a numerical simulation code, a model for this factor ($C$) must be chosen.

\noindent For the $nn$ and $pp$ the momentum-dependent factors
$M_{NN}$ read

$$ M_{nn}= \left(\frac{{\bf |k|}^2}{{\bf |k|}^2+m_{\pi}^2}\right)^2 +
\left(\frac{{\bf |l|}^2}{{\bf |l|}^2+m_{\pi}^2}\right)^2 +\frac{(1-\beta){\bf |k|}^2 \cdot{\bf |l|}^2}{({\bf |k|}^2 +
m_{\pi}^2)({\bf |l|}^2 + m_{\pi}^2)} \eqno(3) $$
with $\beta=3\langle(\textbf{\underline{k}} \cdot \textbf{\underline{l}})^{2}\rangle$ ($\textbf{\underline{k}},\textbf{\underline{l}}$ being the coresponding unit vectors for \textbf{k} and \textbf{l}),

\noindent while for the $np$ process
$$ M_{np} =
 \left(\frac{{\bf |k|}^2}{{\bf |k|}^2+m_{\pi}^2}\right)^2 + 4\left(
\frac{{\bf |l|}^2}{{\bf |l|}^2 + m_{\pi}^2}\right)^2 +2(1-\beta)
\frac{{\bf |k|}^2 \cdot{\bf |l|}^2}{({\bf |k|}^2+ m_{\pi}^2)({\bf
|l|}^2 + m_{\pi}^2)} \eqno(4) $$

\noindent where ${\bf k} = {\bf p}_1 - {\bf p}_3 $ and ${\bf l} = {\bf
p}_1 - {\bf p}_4 $ are the nucleon direct and exchange transfer
momenta, respectively. The last (exchange) terms in the above
expressions arise from interference of two different reaction
amplitudes. They contain contributions from the scalar product
$({\bf k\cdot l})^2$, which have been estimated (\cite{[YAK00]},\cite{[BKT88]}) by replacing
them by their average values (denoted by $\beta$) in the
phase-space. There are two numerical values for $\beta$ in the literature: $\beta=1.0845$, in Ref. \cite{[BKT88]} and $\beta=1.3078$ in Ref. \cite{[GRS95]}, but this difference (explained by Raffelt and Seckel) produces changes in our final results of only 1\%. Since we compare our results especially with those of Ref. \cite{[BKT88]}, we use that value for $\beta$. Thus, from kinematical constraints $\beta$ =0 in the D
regime, while it is $1.0845$ in the ND regime (see the expression above). We have
used for the $np$ process, the NME of \cite{[BKT88]}, with equal coupling constants for protons and neutrons.
%The choice of either set of values of $\xi$ and
%$\zeta$ results in significant changes in the high-momentum limits
%of the NME. %However, our calculations show that Yakovlev's expression for the OPEP in the $np$ case is the right one and gives physical results. Thus, we use this expression for our numerical results.

We follow the procedure of Brinkmann and Turner \cite{[BKT88]} to
derive the ND limit, by performing the transformation to the
center-of-mass system
$$ {\bf p}_+ = \frac{{\bf p}_1 + {\bf p}_2}{2};~~{\bf p}_- = \frac{{\bf p}_1 -
{\bf p}_2}{2};~~{\bf p}_{3c} = {\bf p}_3 - {\bf p}_+;~~{\bf p}_{4c} =
{\bf p}_4 -{\bf p}_+ $$
$$\Rightarrow
{\bf p}_1={\bf p}_+ +{\bf p}_-;~~{\bf p}_2={\bf p}_+-{\bf p}_-;~~
{\bf p}_3={\bf p}_+ +{\bf p}_{3c};~~{\bf p}_4= {\bf p}_++{\bf
p}_{4c} \eqno(5) $$
From these relations and the conservation of momentum (axion momentum is neglected) we find
${\bf p}_{4c} =- {\bf p}_{3c}$. We define also the dimensionless quantities

$$ u_i = \frac{{\bf p}_i^2}{2mT} (i=\bar{1,4});~~u_+=\frac{{\bf p}_+^2}{2mT};~~
u_-=\frac{{\bf p}_-^2}{2mT};~~u_{3c}=\frac{{\bf p}_{3c}^2}{2mT}, \eqno(6)$$
$$ cos{\gamma_1} = \frac{{\bf p}_+ {\bf p}_-}{\vert {\bf p}_+\vert \vert
{\bf p}_-\vert};~~cos{\gamma_c} = \frac{{\bf p}_+ {\bf p}_{3c}}{\vert
{\bf p}_+ \vert \vert {\bf p}_{3c}\vert};~~ cos{\gamma} = \frac{{\bf
p}_- {\bf p}_{3c}}{\vert {\bf p}_-\vert \vert {\bf
p}_{3c}\vert};\eqno(7) $$

From the definition of the $u$ variables above, and the
conservation of energy, one can easily deduce the following
relations
$$ u_{1,2} = u_+ + u_- \pm 2(u_+ u_-)^{1/2}cos{\gamma_1};~~ u_{3,4} = u_+ + u_{3c}
 \pm 2(u_+u_{3c})^{1/2}cos{\gamma_c};$$ $$ u_-=u_{3c}+{E_a}/{2T} \eqno(8) $$

Let us now address the OPE potential. Following the method used in
our previous papers (\cite{[SSH02]},\cite{[SPN04]}), and after some lengthy algebra,
one can express the  matrix element  $M_{nn}$ (eq.(4)) in terms of
the scalar combinations ${\bf |k|}^2 + {\bf |l|}^2$ and ${\bf
|k|}^2 \cdot {\bf |l|}^2$. Finally we expressed these NME in the
following compact form

$$ {S}\times \Sigma {\vert {M}\vert^2} = \frac{64 m^2
g_{ai}^2}{3}\left(\frac{f}{m_{\pi}}\right)^4 \left[(3-\beta) -
|M_{nn}|^2_{nucl}\right] \eqno(9) $$

\noindent
where

$$ |M_{nn}|^2_{nucl} = m^2_{\pi} \frac{A_{nn} - B_{nn}\cdot C_{\phi}^2}
{C - D\cdot C_{\phi}^2 - E \cdot C_{\phi}^4} \eqno(10) $$

The coefficients $A_{nn}, B_{nn}, C, D$ and $E$ of eq.(10) are
polynomials depending on the parameters $m$, $T$ and $m_{\pi}$ and
of variables $u_-$ and $u_{3c}$ (for their full expressions, see
Appendix A). Also we used the notation $C_{\phi}=cos \gamma_1 cos
\gamma_c + sin \gamma_1 sin \gamma_c cos\phi$, with $\phi$ the
angle between the vectors ${\bf p}_+$ and ${\bf p}_-$.

Thus, the contribution of NME is split into a constant
term,obtained also by Brinkmann and Turner (\cite{[BKT88]}), Raffelt and Seckel(\cite{[GRS95]})) - which represents just its high-momentum limit
(i.e. the limit to which the expression (3) converges when the
pion mass is neglected compared to the nucleon momentum transfer)
and a reduction term (see Appendix A) to be evaluated.
After an approximation which is
numerically accurate within $ 1\%$, we succeeded to
integrate the expression of $|M_{nn}|^2$ over the angles and
finally we could express the axion emission rate in the ND limit
in the following form

$$ \epsilon^{ND}_{aNN} = \epsilon^{ND}_{aNN}(0)\left(1-
\frac{I^{ND}_{nucl}(NN)}{(3-\beta)I^{ND}_0}\right) \eqno(11) $$

where
$$\epsilon^{ND}_{aNN}(0)= 2.68 \times
10^{-4}g^2_{ai} e^{2y} m^{2.5} T^{6.5}(f/m_{\pi})^4 \eqno(12)$$
is the expression calculated by Brinkmann and Turner \cite{[BKT88]} and
$I^{ND}_{0}$ and $I^{ND}_{nucl}(NN)$ are double integrals over $u_-$ and
$u_{3c}$

$$ I^{ND}_0 = \int_0^{\infty} \int_0^{u_-}{\sqrt{(u_-u_{3c})}(u_{-}- u_{3c})^2 e^{-2u_-}du_-
du_{3c}} \eqno(13) $$

 $$ I^{ND}_{nucl}(NN) = \frac{\pi m_{\pi}^2}{mT}\int_0^{\infty} \int_0^{u_-}
\sqrt{(u_-u_{3c})}(u_- - u_{3c})^2 e^{-2u_-} $$
$$\times \left( \frac{(7-\beta)m_{1} + 4(3-\beta)(u_- + u_{3c})}{(2u_- + 2u_{3c} + m_1)^2} \right)du_- du_{3c}
\eqno(14) $$

\noindent with $N=n,p$, $m_1 = m_{\pi}^2/mT$.

A similar procedure for the $np$ process yields

$$ {S}\times \Sigma {\vert {M}\vert^2} = \frac{256 m^2
g_{aN}^2}{3}\left(\frac{f}{m_{\pi}}\right)^4 \left[(7-2\beta)
- |M_{np}|^2_{nucl}\right] \eqno(15) $$ where 
$$ |M_{np}|^2_{nucl} = m^2_{\pi} \frac{A_{np} - C_{np}\cdot C_{\phi}^2}
{C - D \cdot C_{\phi}^2 - E \cdot C_{\phi}^4}+C_{\phi} \frac{B_{np} -
D_{np}\cdot C_{\phi}^2} {C - D \cdot C_{\phi}^2 - E \cdot C_{\phi}^4} \eqno(16) $$
and $g_{aN}=[(7-2\beta)/3]g_{ai}^{2}$ is the effective axion nucleon coupling for the np case (see expression A.1 in ref.\cite{[BKT88]}),

and final expressions analogous to Eq.(11) and (12):

$$ \epsilon^{ND}_{anp} = \epsilon^{ND}_{anp}(0)\left(1 -
\frac{I^{ND}_{nucl}(np)}{(7-2\beta)I^{ND}_0}\right) \eqno(17) $$

$$\epsilon^{ND}_{anp}(0)= 2.68 \times
10^{-4}g^2_{aN} e^{y_{1}+y_{2}} m^{2.5} T^{6.5}(f/m_{\pi})^4 \eqno(18)$$

The correction integral of the third term $I^{ND}_{nucl}(NN)$ is replaced by

$$ I^{ND}_{nucl}(np) = \frac{\pi m_{\pi}^2}{mT}\int_0^{\infty} \int_0^{u_-}
\sqrt{(u_-u_{3c})}(u_- - u_{3c})^2 e^{-2u_-}\times $$
$$ \left(\frac{4(7-2\beta)(u_-+u_{3c}) + (17-2\beta)m_1}{(2u_- + 2u_{3c} +
m_1)^2} \right)du_- du_{3c} \eqno(19) $$

Using the same procedure, we calculated the emissivities for the
$nn$($pp$) processes in the D limit. In this case, in performing
the integrals over the momenta and energies, we used the method of
integration adopted in Refs.\cite{[IWA84]},\cite{[IWA01]} and
\cite{[FSB75]}. The expressions of the emissivities are

$$ \epsilon^{D}_{aNN} = \epsilon^{D}_{aNN}(0) \left(1 -
\frac{I^{D}_{nucl}(NN)}{3I^{D}_0}\right) \eqno(20) $$

\noindent where

$$\epsilon^{D}_{aNN}(0)=
\left(\frac{31}{3780\pi}\right)\left(\frac{g^2_{ai}}{\hbar^5
c^7}\right)\left(\frac{f}{m_{\pi}}\right)^4 m^2_n p_F(N)(kT)^6$$
or $$\epsilon^{D}_{aNN}(0)=\frac{31\sqrt{2}}{3780\pi}m^{2.5}T^{6.5}m_{\pi}^{-4}g_{ai}^{2}f^{4}y^{1/2} \eqno(21)$$ (in natural units and $p_{F}(N)=2mTy$)

\noindent
with N = $n$ or $p$, and

$$I^{D}_0 = \int_0^{\infty}{z^3 \left(\frac{4\pi^2 + z^2}{e^z - 1}\right)dxdz}
\eqno(22)$$

$$ I^{D}_{nucl}(NN) = \frac{\pi m_{\pi}^2}{mT}\int_0^{\infty} \int_0^{x_f}
{z^3 \left(\frac{4\pi^2 + z^2}{e^z - 1}\right)\times\left(\frac{6(2x+z)
+ 7m_1}{(2x+z+m_1)^2}\right)dxdz} \eqno(23) $$

Similar calculations for the $np$ process yields a final expression analogous to Eq. (17):

$$ \epsilon^{D}_{anp} = \epsilon^{D}_{anp}(0) \left(1 -
\frac{I^{D}_{nucl}(np)}{7I^{D}_0}\right) \eqno(24) $$

with

$$\epsilon^{D}_{anp}(0)=\frac{31\sqrt{2}}{3780\pi}m^{2.5}T^{6.5}
m_{\pi}^{-4}g_{aN}^{2}f^{4}y_{m}^{1/2}(1-\Delta y/2y_{m})
\eqno(25)$$

(expression obtained in ref. \cite{[BKT88]}), where $y_{m}=(y_{1}+y_{2})/2$, $\Delta y=\vert{y_{1}-y_{2}}\vert/2 $,

and the correction integral
$$ I^{D}_{nucl}(np)= {\frac{\pi m_{\pi}^{2}}{mT}\int_{0}^{\infty}
\int_{0}^{x_{f}}z^{3}\left(\frac{4\pi^{2}+z^{2}}{e^{z}-1}\right)\times}$$
$${\times\left(\frac{14(2x+z)+17m_{1}}{(2x+z+m_{1})^{2}}\right)dxdz} \eqno(26)$$

With these expressions at hand, we discuss the results in next
section.

\section{Results}

In Figs. 1 and 2 we plotted the dependence on temperature of the relative corrections of the emissivities, $\Delta\epsilon/\epsilon_{0}=(\epsilon_{0}-\epsilon)/\epsilon_{0}$ (where $\epsilon_{0}$ is the emissivity calculated in the high momentum limit - see Ref. \cite{[BKT88]}, while $\epsilon$ is the emissivity determined with our method), for ND regime (Fig. 1 - $nn/pp$ processes and Fig.2 - $np$ process) due to pion mass effects (see Eqs. (11), (17)). One observes that for this regime, the relative corrections to the previous results for the emissivities are quite important and temperature dependent. For all processes, the corrected emissivities are reduced with 30\% to 85\%, depending on temperature.
Figs. 3 and 4 show the same relative emissivities, this time for the D regime (Fig. 3 - $nn/pp$ processes and Fig.4 - $np$ process). In this case, the corrected emissivities are reduced at most with 11\%. So, for this regime the contributions from pion mass effects (see Eqs. (20),(24)) do not seriously affect the corresponding emission rates (a result qualitatively reported in \cite{[GRS95]}).The dependence of these corrections on temperature is smoother in this case than for the ND regime.
For more confidence, we give in Table 1 the absolute values for $\Delta\epsilon/\epsilon_{0}$ for ND and D regimes. We remark a very similar behaviour of the $nn/pp$ and $np$ relative corrections, for both regimes.

\begin{table}[]
 \begin{tabular}{lccccl}
\hline
$T[MeV]$ & $(\Delta\epsilon/\epsilon_{0})_{ND}^{nn/pp}$ &  $(\Delta\epsilon/\epsilon_{0})_{ND}^{np}$ & $T[MeV]$ & $(\Delta\epsilon/\epsilon_{0})_{D}^{nn/pp}$ &  $(\Delta\epsilon/\epsilon_{0})_{D}^{np}$ \endline  
\hline
$25$ & $0.8503$ & $0.8490$ & $1$ & $0.1030$ & $0.1070$ \endline 
$30$ & $0.7189$ & $0.7180$ & $2$ & $0.0960$ & $0.0997$ \endline
$35$ & $0.6234$ & $0.6227$ & $3$ & $0.0900$ & $0.0931$ \endline
$40$ & $0.5503$ & $0.5497$ & $4$ & $0.0846$ & $0.0874$ \endline
$45$ & $0.4923$ & $0.4918$ & $5$ & $0.0800$ & $0.0824$ \endline
$50$ & $0.4455$ & $0.4450$ & $6$ & $0.0756$ & $0.0780$ \endline
$55$ & $0.4070$ & $0.4066$ & $7$ & $0.0720$ & $0.0740$ \endline
$60$ & $0.3738$ & $0.3735$ & $8$ & $0.0686$ & $0.0705$ \endline
$65$ & $0.3459$ & $0.3455$ & $9$ & $0.0653$ & $0.0673$ \endline
$70$ & $0.3224$ & $0.3220$ & $10$ & $0.0627$ & $0.0643$ \endline
$75$ & $0.3010$ & $0.3008$ & $20$ & $0.0440$ & $0.0435$ \endline
\hline
\end{tabular}
\caption{Relative correction to the emissivities (relative emissivities) for ND and D regime, for all NN bremsstrahlung processes ($nn,pp,np$). Here $\Delta\epsilon/\epsilon_{0}=(\epsilon_{0}-\epsilon)/\epsilon_{0}$, with $\epsilon$ - the emission rates determined with our method and $\epsilon_{0}$ - the emissivities previously obtained in Ref. \cite{[BKT88]}} 
\end{table}

\begin{figure}[]

\includegraphics[height=0.40\textheight,width=0.80\textwidth]{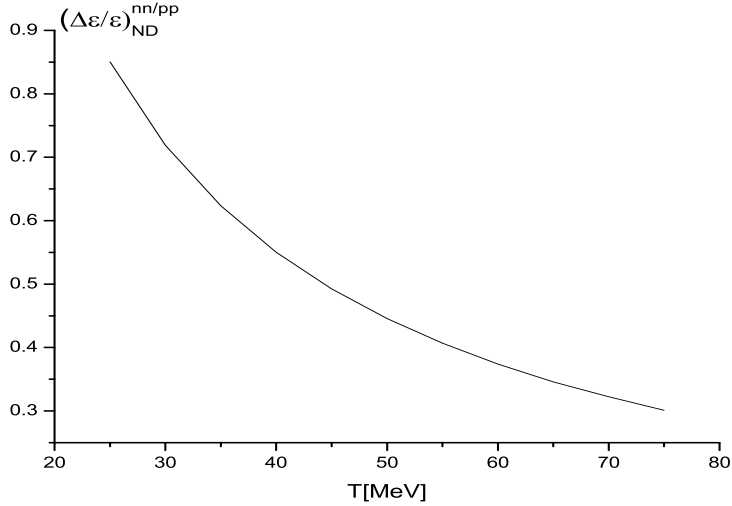}
\caption{\label{fig:Graph17}Relative correction to the emissivities (Relative emissivities) due to pion mass effects for the ND regime, $nn,pp$ processes.}

\end{figure}

%\begin{table}[]
%\begin{tabular}{lcccl}
%\hline
%$T$ [MeV] & $\epsilon_{aNN}^{ND}(y=0)$ &  $\epsilon_{aNN}^{ND}(y=-1)$ &  $\epsilon_{aNN}^{ND}(y=-2)$ \endline
%\hline
%$25$ & $3.222*10^{33}$ & $4.359*10^{32}$ & $5.897*10^{31}$ \endline
%$35$ & $3.235*10^{34}$ & $4.378*10^{33}$ & $5.923*10^{32}$ \endline
%$45$ & $1.763*10^{35}$ & $2.386*10^{34}$ & $3.229*10^{33}$ \endline
%$55$ & $6.743*10^{35}$ & $9.124*10^{34}$ & $1.234*10^{34}$ \endline
%$65$ & $2.049*10^{36}$ & $2.773*10^{35}$ & $3.752*10^{34}$ \endline
%$75$ & $5.292*10^{36}$ & $7.163*10^{35}$ & $9.694*10^{34}$ \endline
%\hline
%$T$ [MeV] & $\epsilon_{aNN}^{ND}(y=-3)$ &  $\epsilon_{aNN}^{ND}(y=-4)$ &  $\epsilon_{aNN}^{ND}(y=-10)$ \endline
%\hline
%$25$ & $7.980*10^{30}$ & $1.080*10^{30}$ & $6.647*10^{24}$ \endline
%$35$ & $8.017*10^{31}$ & $1.085*10^{31}$ & $6.677*10^{25}$ \endline
%$45$ & $4.370*10^{32}$ & $5.915*10^{31}$ & $3.640*10^{26}$ \endline
%$55$ & $1.670*10^{33}$ & $2.260*10^{32}$ & $1.391*10^{27}$ \endline
%$65$ & $5.077*10^{33}$ & $6.872*10^{32}$ & $4.229*10^{27}$ \endline
%$75$ & $1.320*10^{34}$ & $1.786*10^{33}$ & $1.099*10^{28}$ \endline
%\hline
%\end{tabular}
%\caption{Axion-bremsstrahlung emissivities for $np$ process
%$(\xi=2,\zeta=-2(1-\beta/3)=-1.277)$, in ND case. As before, the
%emissivities are divided by $C^{2}$ and measured in $erg \, cm^{-3} \, s^{-1}$.}
%\end{table}

\begin{figure}[]

\includegraphics[height=0.40\textheight,width=0.80\textwidth]{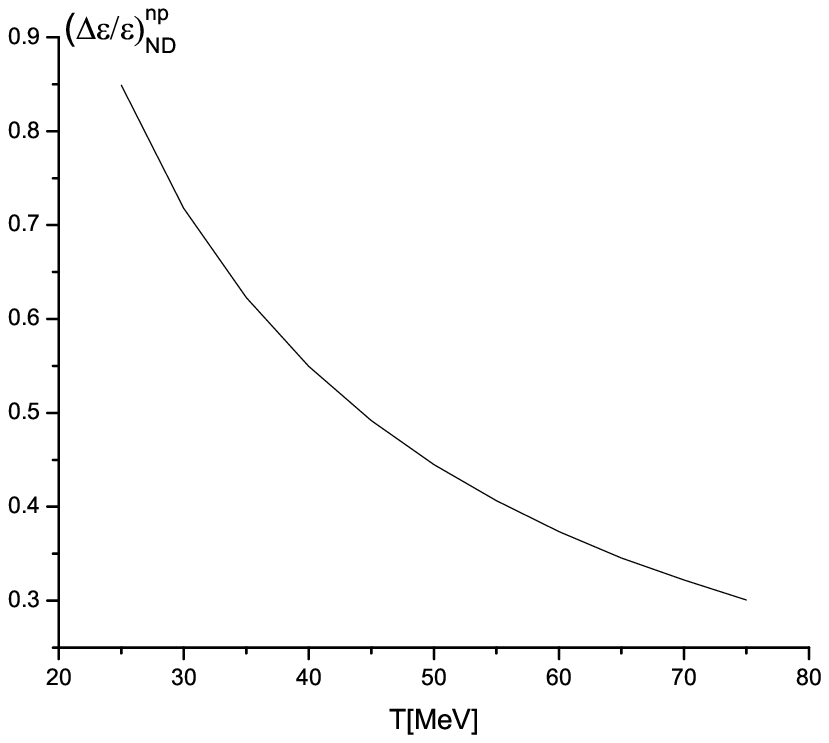}
\caption{\label{fig:Graph18}Relative emissivities due to pion mass effects for the ND regime, $np$ process.}

\end{figure}

\begin{figure}[]

\includegraphics[height=0.40\textheight,width=0.80\textwidth]{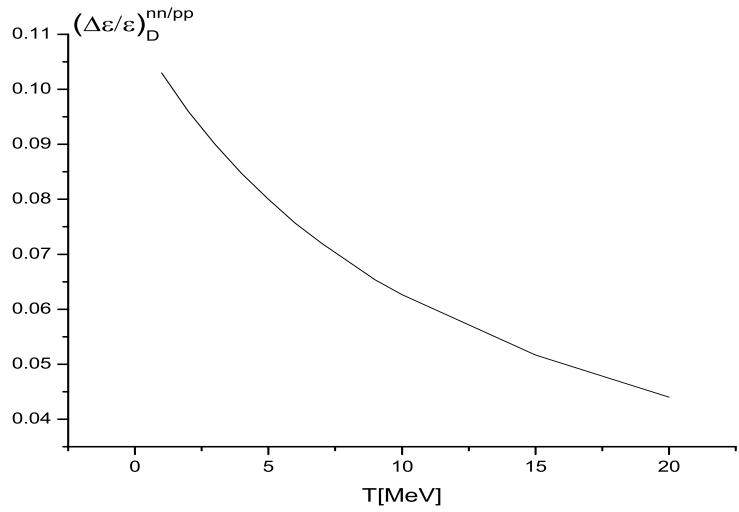}
\caption{\label{fig:Graph19} Relative correction to the emissivities due to pion mass effects for the D regime, $nn,pp$ processes.}

\end{figure}

\begin{figure}[]

\includegraphics[height=0.40\textheight,width=0.80\textwidth]{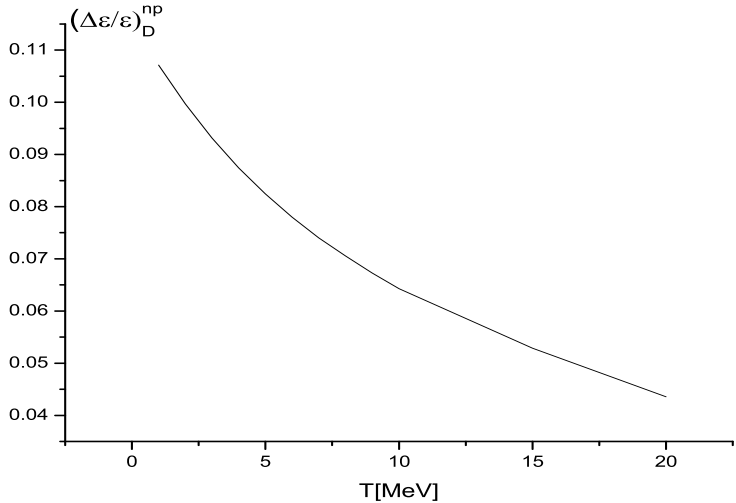}
\caption{\label{fig:Graph20}Relative emissivities due to pion mass effects for the ND regime, $np$ process.}

\end{figure}

\newpage

\section{Conclusions}

In this paper we presented a method of determination of the NN
axion-bremsstrahlung emissivities, for all processes, in both ND
and D regimes, based on the inclusion of the full dependence of the NME on nuclear momenta.

Starting with the ND case, worked out for the $nn$, $pp$ and $np$
cases, we found substantial reductions to the emissivities due to
the combined effects of momentum-dependence in the
temperature-dependent integrals - Eqs. (14),(19). While these
effects were somewhat foreseen in previous works (\cite{[BKT88]}), here
we present an explicit calculation and quantitative results. We
claim, based on the present results, that all axion mass limits
that employed $\epsilon_{aNN}$ should be revised (the ones used in numerical simulation supernova codes), especially those
for which the ND limit is highly relevant, such as SN1987A
neutrinos (\cite{[YAK00]}). As it
stands, the suppression of $\epsilon_{aNN}$ can be important and
this feature postpones one to obtain a firm bound on $m_{a}$
for both popular KSVZ (\cite{[KIM79]},\cite{[SVZ80]}) and DSVZ
(\cite{[DFS81]},\cite{[AZH80]}) axion models.

For the D case, our method allowed to check that the leading terms
(Eq. (21), (25)) coincide with Iwamoto
(\cite{[IWA84]},\cite{[IWA01]}) and Brinkmann and Turner
(\cite{[BKT88]}) results (see for example Eqs.(2.6) and (2.8) of
Ref. \cite{[IWA01]}), and also that the reductions are up to 11\% for all three processes (a-c).
 %(as already described by Iwamoto's formulae) but also mildly temperature-dependent, as expected from physical considerations.

We also mention that other nuclear effects like spin-density fluctuations (\cite{[WJS97]},\cite{[GSI96]}) or short range effects (TPEP) can further reduce the coresponding axion emissivities.We are currently calculating the NN axion-bremsstrahlung emissivities by including two pion exchange effects (through a TPEP) that can be mimicked by the exchange of a $\rho$ meson (\cite{[ERM89]}). These effects are important at distances below 2 fm (\cite{[ERW88]}). Preliminary results show us a further reduction of the axion emission rates compared with the case presented here. The calculations will be reported in a future paper. For the moment it is fair to point out that apparently minor sources of error become actually important for the problem, already at the OPE
approximation level, as explicitly shown above. Also, it is worth to mention that our method might be used to improve calculations for other physical processes of neutrino and axion emission in NS, in which strong interactions are also involved.

\textbf{Acknowledgments}

S.S and B.P. acknowledge support from CNCSIS, Project ID-975, Contract 
No. 563/14.01.2009.
J.E.H. and M.P. Allen wish to thank CNPq and Fapesp Agencies (Brazil) 
for financial support through grants and fellowships.

\appendix
\section{Appendix A}

We present here the correction matrix elements for all the cases
and processes, which have not been presented in section 2 (Calculations). Starting with ND regime, for nn and pp processes, the
correction matrix element is

$$ |M_{nn}|^2_{nucl} = m^2_{\pi} \frac{A_{nn} - B_{nn}\cdot C_{\phi}^2}
{C - D\cdot C_{\phi}^2 - E \cdot C_{\phi}^4} \eqno(A.1) $$

where
$$ A_{nn}= 2(3-\beta)m_{t}^{3}U_{+}^{3}+5(3-\beta)m_{\pi}^{2}m_{t}^{2}U_{+}^{2}+4(3-\beta)m_{\pi}^{4}m_{t}U_{+} \eqno(A.2)$$
$$ B_{nn}=4m_{t}^{2}U_{p}[2(3-\beta)m_{t}U_{+}+(7-\beta)m_{\pi}^{2}]  \eqno(A.3) $$
$$ C=m_{t}^{4}U_{+}^{4}+4m_{t}^{3}m_{\pi}^{2}U_{+}^{3}+6m_{\pi}^{4}m_{t}^{2}U_{+}^{2}
+4m_{\pi}^{6}m_{t}U_{+}+m_{\pi}^{8}       \eqno(A.4)$$
$$ D= 8m_{t}^{2}U_{p}(m_{t}U_{+}+m_{\pi}^{2})^{2}  \eqno(A.5) $$
$$ E=16m_{t}^{4}U_{p}  \eqno(A.6) $$  
with  $$m_{t}=2mT; U_{+}=u_{-}+u_{3c}; U_{p}=u_{-}u_{3c} \eqno(A.7) $$

In ND case, for the np process we obtained the correction matrix
elements and the corresponding coefficients (C, D and E are the
same) as follows:

$$ |M_{np}|^2_{nucl} = m^2_{\pi} \frac{A_{np} - C_{np}\cdot C_{\phi}^2}
{C - D \cdot C_{\phi}^2 - E \cdot C_{\phi}^4}+ \frac{B_{np}C_{\phi} -
D_{np}\cdot C_{\phi}^3} {C - D \cdot C_{\phi}^2 - E \cdot C_{\phi}^4} \eqno(A.8)$$

where
$$ A_{np}=(7-2\beta)[2m_{t}^{3}U_{+}^{3}+5m_{\pi}^{2}m_{t}^{2}U_{+}^{2}+m_{\pi}^{4}(m_{\pi}^{2}+4m_{t}U_{+})] \eqno(A.9)$$
$$ B_{np}=12m_{t}^{2}U_{+}U_{p}^{1/2}(m_{t}U_{+}+m_{\pi}^{2}) \eqno(A.10) $$
$$ C_{np}=4[2(7-2\beta)m_{t}^{3}U_{+}U_{p}+(17-2\beta)m_{t}^{2}m_{\pi}^{2}U_{p}] \eqno(A.11) $$
$$ D_{np}=48m_{t}^{3}U_{p}^{3/2} \eqno(A.12) $$

For the D regime, the expressions for the correction matrix
elements and for the corresponding coefficients are obtained by
taking $\beta=0$ in previous relations (A.1-A.12).

\end{document}